\documentclass[runningheads]{llncs}
\usepackage[T1]{fontenc}
\usepackage{amsmath}
\usepackage{latexsym}
\usepackage{multirow}
\usepackage{color}
\usepackage[usenames,dvipsnames,table]{xcolor}
\usepackage{graphicx}
\usepackage{epsfig} 
\usepackage{epsf}   
\usepackage{dcolumn}
\usepackage{bm}
\usepackage{dcolumn}
\usepackage{textcomp}
\usepackage{float}
\usepackage{subfig}
\usepackage{makecell}
\usepackage{color}
\usepackage{pifont}
\usepackage{appendix}
\usepackage{multirow,bigdelim}  
\usepackage{lineno}

\usepackage{hyperref}
\usepackage[numbers,sort&compress]{natbib}

\usepackage[compat=1.0.0]{tikz-feynman}

\usepackage{graphicx}
\begin{document}
\title{Neutrino mass ordering sensitivities at DUNE, HK and KNO in presence of scalar NSI \\ \vspace{0.6cm} \small Contribution to the 25th International Workshop on Neutrinos from Accelerators}
\titlerunning{Neutrino MO sensitivities at DUNE, HK and KNO in presence of SNSI}
%

\author{Moon Moon Devi\inst{1,*} \and
Dharitree Bezboruah\inst{1} \and
Abinash Medhi\inst{2} \and
Arnab Sarker\inst{1}\and
Debajyoti Dutta\inst{3}}
\authorrunning{M.M.Devi et al.}

%
\institute{Department of Physics, Tezpur University, Napaam, Assam, India \and
IIT Guwahati, Assam, India \and
Bhattadev University, Bajali, Assam, India
\\
\email{Email: $^{*}$devimm@tezu.ernet.in}}
\maketitle              
\begin{abstract}
The limitations of the Standard Model in explaining neutrino masses and neutrino mixing leads to the exploration of frameworks beyond the Standard Model (BSM). The possibility of neutrinos interacting with fermions via a scalar mediator is one of the interesting prospects. The study of neutrino non-standard interactions (NSI) is a well-motivated phenomenological scenario to explore new physics beyond the Standard Model. These new interactions may alter the standard neutrino
oscillation probabilities, potentially leading to observable effects in experiments. It also allows for the exploration of absolute neutrino masses via oscillation experiments. It can modify the oscillation probabilities, which in turn can affect the physics sensitivities in long-baseline experiments. The linear scaling of the effects of scalar NSI with matter density also motivates its exploration in long-baseline (LBL) experiments. We will present our study on the impact of a scalar-mediated NSI on the mass ordering (MO) sensitivities of three long-baseline neutrino experiments, i.e., DUNE, HK and KNO. We study the impact on MO sensitivities at these experiments assuming that scalar NSI parameters are present in nature and are known from other non-LBL experiments. The presence of scalar NSI can notably impact the MO sensitivities of these experiments. Furthermore, we analyze the potential synergy by combining data from DUNE with HK and HK+KNO, thereby exploring a broader parameter space.

\end{abstract}

\section{Introduction}
The phenomenon of neutrino oscillations provides the first experimental proof of physics beyond the standard model(BSM). Several outstanding experiments worldwide are now aimed at improving the accuracy of the measurements of the mixing parameters, determination of the mass ordering and the leptonic CP violation. The BSM models for neutrino mass and mixing inherently suggest new particles and interactions. The scalar non-standard interaction (SNSI) is one such interesting BSM phenomenon. Scalar NSI \cite{Ge:2018uhz,Babu:2019iml} is the neutrino-fermion interaction mediated by a scalar particle. The idea of such non-standard interactions (NSIs) was first proposed by Wolfenstein\cite{Wolfenstein:1977ue} as a possible explanation of flavor change when neutrinos propagate. It is now established that flavor change is primarily due to oscillation, and NSIs act as subdominant matter effects. These subdominant effects can significantly impact the precision measurement of $\nu$-oscillation parameters in present and future high-precision experiments. In this work, we explore the mass ordering (MO) sensitivities of DUNE \cite{DUNE:2020txw} and HK+KNO\cite{Hyper-Kamiokande:2016srs}.

\section{Scalar NSI Formalism}
Neutrinos may interact with matter fermions via a scalar mediator and the Lagrangian in the presence of SNSI can be framed as,
\begin{equation}\label{eq:nsi_L}
{\cal L}_{\rm eff}^{\rm S} = \frac{y_f y_{\alpha\beta}}{m_\phi^2}(\bar{\nu}_\alpha(p_3) \nu_\beta(p_2))(\bar{f}(p_1)f(p_4)),
\end{equation}
where, $\alpha$, $\beta$ refer to the neutrino flavors (e, $\mu$, $\tau$); and $f = e, u, d$ indicate the matter fermions, ($e$: electron, $u$:up-quark, $d$: down-quark), while, $\bar{f}$ indicates the corresponding anti-fermions, $y_{\alpha\beta}$ is the Yukawa couplings of the $\nu's$ with $\phi$, $y_f$ is the Yukawa coupling of $\phi$ with $f$, and $m_\phi$ is the mass of $\phi$. The SNSI contribution appears as a medium-dependent perturbation to the neutrino mass term as the Yukuwa term in the Lagrangian constrains us from converting it to vector currents. Hence, the effective form of the Hamiltonian in presence of SNSI is given as,
\vspace{-5pt}
\begin{equation}
\mathcal H_{\rm SNSI}
\equiv
E_\nu
+ \frac { M_{\rm eff}  M_{\rm eff}^\dagger}{2 E_\nu}
\pm V_{\rm SI} \,,
\label{eq:Hs}
\end{equation}

where, $M_{\rm eff}$ = $M + \delta M$, is the effective $\nu$-mass matrix that includes both the regular mass matrix $M$ and the contribution from the SNSI, $\delta M  \equiv \sum_f n_f y_f y_{\alpha\beta} / m^2_\phi$. We have parameterized the scalar NSI contribution as in references \cite{Ge:2018uhz,Sarker:2023qzp} as,

\begin{equation}
\delta M
\equiv
S_m
\begin{bmatrix}
\eta_{ee}     & \eta_{e \mu}    & \eta_{e \tau}   \\
\eta_{e \mu}^* & \eta_{\mu \mu}  & \eta_{\mu \tau} \\
\eta_{e \tau}^* & \eta_{\mu \tau}^* & \eta_{\tau \tau}
\end{bmatrix}.
\label{eq:dM}
\end{equation}
Here, $S_m$ is the constant scaling term with the dimension of mass. We have used $S_m = \sqrt{2.5 \times 10^{-3}~{\rm eV^2}} \approx 0.05~{\rm eV}$, which corresponds to a typical value of $\Delta m_{31}^{2}$. The dimensionless elements $\eta_{\alpha \beta}$ quantifies the strength of SNSI. 

\section{Methodology}
In this work, we have explored the MO sensitivities of LBL experiments i.e. DUNE, and HK+KNO in the presence of diagonal SNSI parameters. For the sensitivity studies, we have used the GLoBES \cite{Huber:2004ka} framework and used the benchmark values of the oscillation parameters as listed in the table \ref{tab:mixing_parameters}.
\vspace{-10pt}
 \begin{table}[!h]
	\centering
	\begin{tabular}{|c|c|c|c|c|}
		\hline
		Parameters & True Values & Parameters & True Values\\
		\hline
		$\theta_{12}$ & 34.51$^\circ$ & $\delta_{CP}$ & -$\pi$/2\\
		$\theta_{13}$ & 8.44$^\circ$ &  $\Delta m_{21}^2$ & 7.56 $\times$ 10$^{-5}$ $eV^2$\\
		$\theta_{23}$ & 47$^\circ$ & $\Delta m_{31}^2$ & 2.43 $\times$ 10$^{-3}$ $eV^2$\\
		\hline
	\end{tabular}
	\caption{Benchmark values of oscillation parameters used in our analysis \cite{Esteban:2024eli}.} \label{tab:mixing_parameters}
\vspace{-20pt}
\end{table} 

\noindent In order to study MO sensitivity, we define a statistical $\chi^2$ as
\begin{equation}\label{eq:chisq}
\chi^2 \equiv  \min_{\eta}  \sum_{i} \sum_{j}
\frac{\left[N_{true}^{i,j} - N_{test}^{i,j} \right]^2 }{N_{true}^{i,j}},
\end{equation}
where, $N_{true}^{i,j}$ ($N_{test}^{i,j}$) are the number of true (test) events in the $\{i,j\}$-th bin. We have considered the higher $\theta_{23}$ octant as the true octant.

\section{Results and Discussions}
\begin{figure}[!h]
\includegraphics[width=0.32\linewidth, height=8cm]{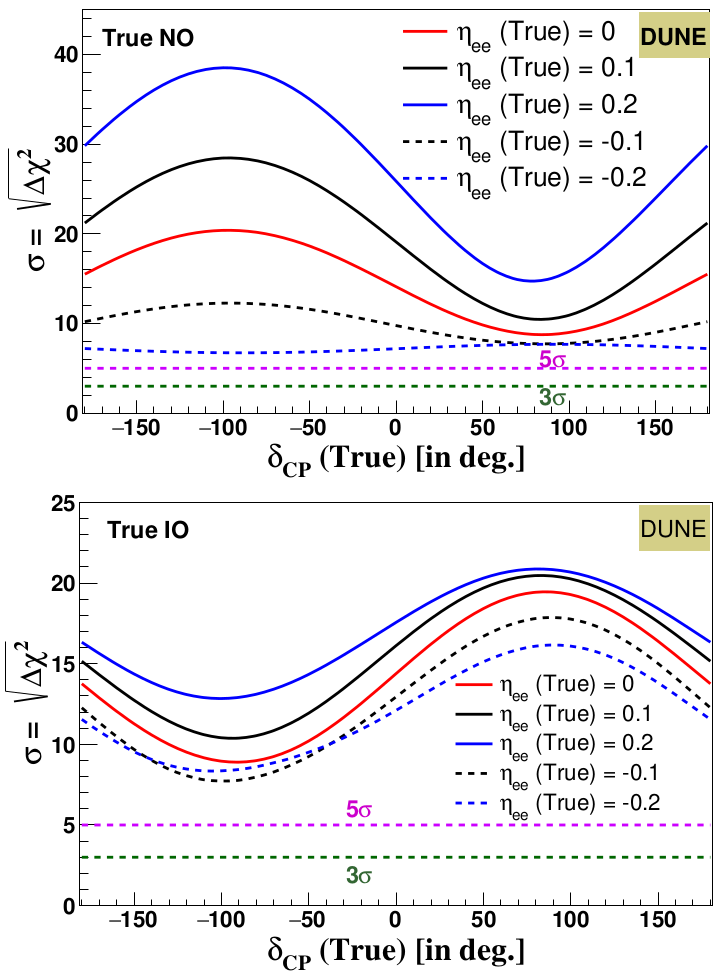}
\includegraphics[width=0.32\linewidth, height=8cm]{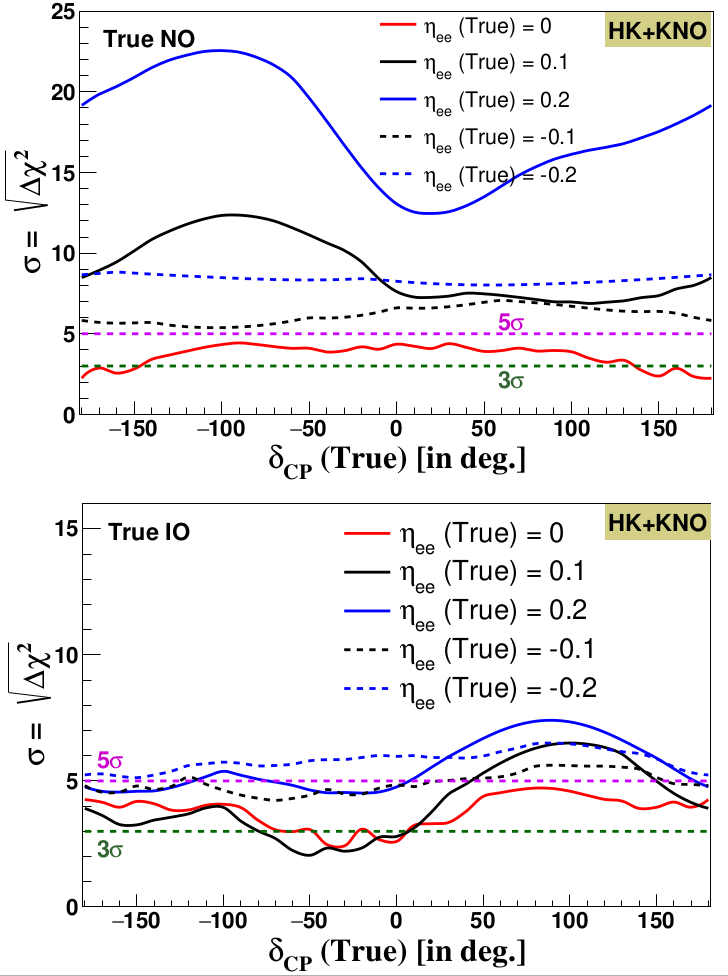}
\includegraphics[width=0.32\linewidth, height=8cm]{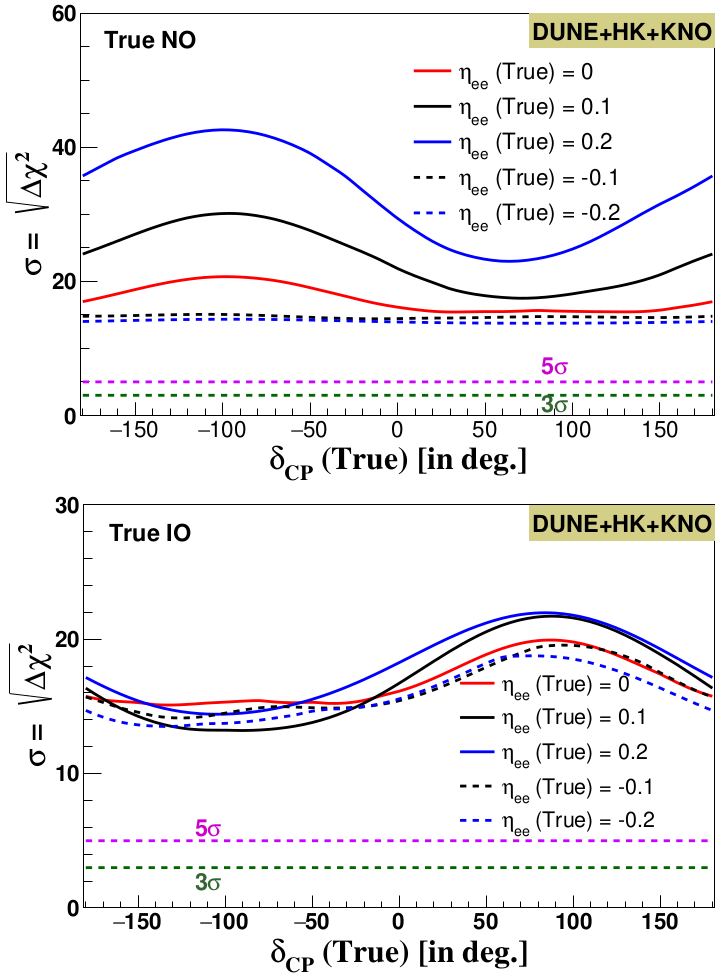}
\caption{MO sensitivities of DUNE (left panel), HK+KNO (middle panel) and DUNE+HK+KNO (right panel) in presence of SNSI element $\eta_{ee}$.} \label{fig:MO_sen}
\end{figure}

\begin{figure}[h!]
\includegraphics[width=\linewidth, height=4cm]{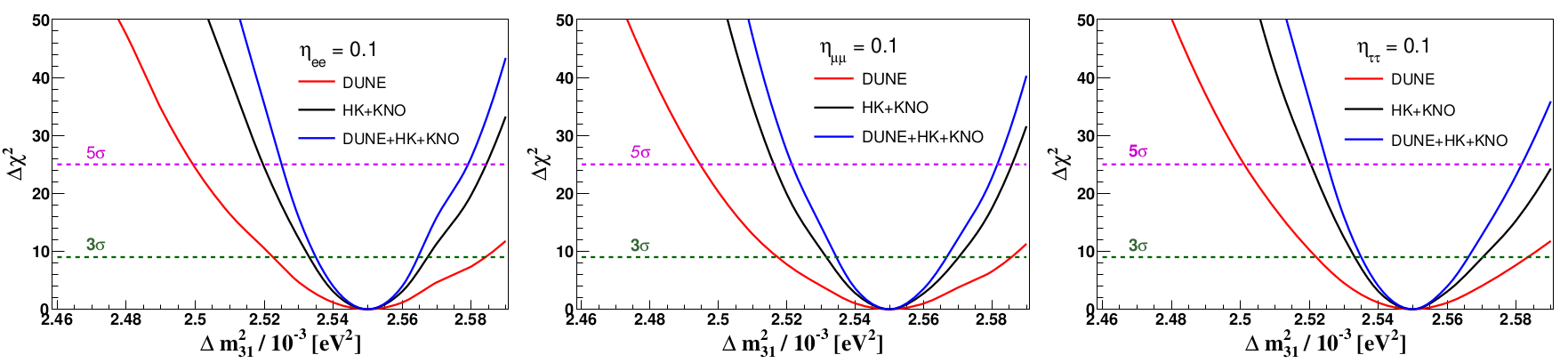}
\caption{The precision measurement of $\rm \Delta m_{31}^2$ in presence of $\eta_{ee}$ (left), $\eta_{\mu \mu}$ (middle) and $\eta_{\tau \tau}$ (right) for DUNE (red), HK+KNO(black) and DUNE+HK+KNO (blue).} \label{fig:MO_prec}
\end{figure}

We have presented the MO sensitivities of DUNE and HK+KNO in the presence of diagonal SNSI element $\eta_{ee}$. We have also explored the MO sensitivities considering the synergy of DUNE and HK+KNO. In figure \ref{fig:MO_sen}, we have shown the MO sensitivities in the presence of $\eta_{ee}$, considering NO (IO) as the true ordering in the upper (lower) panel. The solid (dashed) lines correspond to positive (negative) SNSI parameter values. We have marginalized over $\delta_{CP}$, $\theta_{23}$ and $\eta_{ee}$. We observe that for true NO, a positive (negative) $\eta_{ee}$ enhances (suppresses) the MO sensitivities as compared to the case with no SNSI for DUNE. For true IO scenario, the sensitivities get enhanced (suppressed) as compared to the SI case. However, for HK+KNO the sensitivities depend on the combination of $\delta_{CP}$ and the value of $\eta_{ee}$ for both mass orderings. The synergy of DUNE and HK+KNO leads to an improvement for positive $\eta_{ee}$ in the NO scenario. In figure \ref{fig:MO_prec}, we have explored the precision measurement of $\rm |\Delta m_{31}^{2}|$ for DUNE, HK+KNO and the synergy (DUNE+HK+KNO). In presence of $\eta_{ee}$ (left), $\eta_{\mu\mu}$ (middle) and $\eta_{\tau\tau}$ (right), the $\rm \Delta m_{31}^{2}$ constraining capability of DUNE+HK+KNO configuration is better compared to DUNE and HK+KNO alone. We note that the synergy leads to a better constraining of $\rm |\Delta m_{31}^{2}|$ for all the diagonal elements.

\section{Conclusions}
In this work, we have explored the impact of SNSI on the neutrino MO sensitivities in LBL experiments, particularly DUNE and HK+KNO. We have observed that the presence of $\eta_{ee}$ can significantly affect the MO sensitivities at DUNE and HK+KNO. The synergy of DUNE and HK+KNO also significantly improves the sensitivities due to extended parameter space and higher statistics. The precision measurement of $|\Delta m_{31}^{2}|$ is much better when the synergy of the experiments is considered. 

\subsubsection{Acknowledgements} 
MMD acknowledges SERB, DST for the grant CRG/2021 /002961. AS acknowledges CSIR-HRDG (09/0796(12409)/2021-EMR-I). DB acknowledges DST INSPIRE for the financial support provided through the INSPIRE Fellowship.

%
%
%
 \bibliographystyle{unsrtnat}

\bibliography{main}

\end{document}